\documentclass[a4paper,12pt]{article}
\pdfoutput=1

\usepackage{amsfonts,amsthm}
\usepackage{jheppub}

\newcommand{\mpl}{M_{\rm P}}
\newcommand{\beq}{\begin{equation}}
\newcommand{\eeq}{\end{equation}}
\newcommand{\beaa}{\begin{align}}
\newcommand{\eeaa}{\end{align}}

\title{Pole Inflation - Shift Symmetry and Universal Corrections}

\author[1]{B. J. Broy}
\author[2]{M. Galante}
\author[2]{D. Roest}
\author[1]{A. Westphal}

\affiliation[1]{\emph{Deutsches Elektronen-Synchrotron DESY, Theory Group, 22603 Hamburg, Germany}}
\affiliation[2]{\emph{Van Swinderen Institute for Particle Physics and Gravity, University of Groningen, Nijenborgh 4, 9747 AG Groningen, The Netherlands}}

\emailAdd{benedict.broy@desy.de}
\emailAdd{m.galante@rug.nl}
\emailAdd{d.roest@rug.nl}
\emailAdd{alexander.westphal@desy.de}

\abstract{An appealing explanation for the Planck data is provided by inflationary models with a singular non-canonical kinetic term: a Laurent expansion of the kinetic function translates into a potential with a nearly shift-symmetric plateau in canonical fields. The shift symmetry can be broken at large field values by including higher-order poles, which need to be hierarchically suppressed in order not to spoil the inflationary plateau. The herefrom resulting corrections to the inflationary dynamics and predictions are shown to be universal at lowest order and possibly to induce power loss at large angular scales. At lowest order there are no corrections from a pole of just one order higher and we argue that this phenomenon is related to the well-known extended no-scale structure arising in string theory scenarios. Finally, we outline which other corrections may arise from string loop effects.}

\arxivnumber{1507.02277}

\begin{document}
\maketitle
\flushbottom

\section{Introduction}
 
Recent years have seen increasingly accurate and meaningful observational evidence from probes of the cosmic microwave background (CMB) such as WMAP, Planck, and BICEP2/Keck Array \cite{1212.5225,Ade:2015lrj,Ade:2015tva}, for an early phase of cosmological inflation. Inflation means the existence of a very early phase of typically exponentially accelerated expansion of the universe driven by the large vacuum energy of a scalar field, that terminates in a reheating process which in turn creates the very hot initial phase of conventional big bang cosmology \cite{Guth:1980zm,Linde:1981mu,Albrecht:1982wi}. The simplest models of this process create inflation via vacuum energy domination of a single slowly rolling scalar field. Inflation can produce a large, old, and spatially flat universe if it lasts longer than 60 e-folds of scale factor growth (more unconventional assumptions about the reheating temperature may lower the required amount to $\sim 30$). Moreover, inflation naturally generates a nearly scale-invariant spectrum for the curvature perturbation that is seeding structure formation. The scale of the inflationary scalar potential and its first two derivatives may be chosen in a suitable way such that the three properties reproduce the inflationary observables consistent with experiments.

The arguably simplest models of inflation arise as large-field models where inflation is driven by a single scalar field with a monomial scalar potential \cite{Linde:1983gd}. In this case, justifying the form of the scalar potential over large field displacements requires the presence of a protective weakly broken shift symmetry in the UV completion (see e.g.\ \cite{Kawasaki:2000yn}). Under these assumptions, choosing a trans-Planckian and hence large initial field displacement is sufficient to guarantee enough inflation, and choosing the overall scale of the monomial potential suffices to generate the amplitude of the curvature perturbation in agreement with observations.

All inflation models necessarily generate gravitational waves with a nearly scale-invariant power spectrum in addition to the curvature perturbation. These tensor modes generate a primordial B-mode polarisation pattern in the CMB. The relative strength of this tensor mode power spectrum, the tensor-to-scalar ratio $r$, is controlled by the scale of the inflationary scalar potential and the field range traversed during the observable period of slow-roll inflation via the Lyth bound \cite{Lyth:1996im, Garcia-Bellido:2014wfa}. Large-field models create an observably strong B-mode signal corresponding to $r\gtrsim 0.01$ this way.

However, the data of the recent CMB probes already constrains the tensor mode fraction to $r<0.095$ at $95\,\%$ confidence level \cite{Ade:2015lrj,Ade:2015tva}. Hence, we may wish to construct slow-roll models with smaller field excursion $\Delta\phi\lesssim {\cal O}(a\,few\,\mpl)$ such that they produce $r\lesssim 0.01$ while preserving something akin to the technical naturalness of shift symmetry based large-field models. The prototypical example of such a class of models is the Starobinsky model \cite{Starobinsky}. Its scalar potential becomes exponentially flat with increasing field displacement which manifests a restoration of an approximate shift symmetry. By conformal transformations of the metric one can interpret the Starobinsky scalar as the scalar degree of freedom of an $R+R^2$ model of gravity, or a Jordan frame non-minimally coupled scalar field with approximate scale invariance at larger field values. In this language, recent work established a set of attractor points for various rather general classes of potentials \cite{conformal-inflation, Ferrara:2013, Kallosh:2013tua,Kallosh:2013yoa}, with predictions comparable to those of the Starobinsky model. Yet more recently, it was emphasised that these attractor properties can be rephrased as non-trivial kinetic terms which have a pole of second order in the non-canonical inflaton~\cite{Galante:2014ifa}.

In this paper we expound on these ideas by establishing a duality between a kinetic function with a certain pole structure and shift symmetry of the  Einstein frame canonically normalised inflaton potential. Since non-canonical kinetic terms are a generic consequence of compactifications of higher-dimensional models such as string theory, this may provide a new avenue of constructing this set of phenomenologically promising models from more fundamental embeddings. 

The rest of the paper is structured as follows. First, we elaborate on the importance of approximate shift-symmetries in the context of inflation and describe different manifestations thereof. Then, we recall the formulation of inflation where the inflationary dynamics' complexity has been shifted in parts to the kinetic term rather than the potential. Given a suppression hierarchy for poles of increasing order, we continue to describe corrections to the aforementioned formalism and derive leading order corrections to the inflationary observables. Assuming the corrections to follow the pattern of shift-symmetry in an EFT sense, we then study an infinite tower of corrections and demonstrate that the leading order corrections coincide with the structure obtained before. After outlining phenomenological fingerprints deriving from the corrections, we attempt to embed the previous considerations into some UV theory. Following a generic argument as to what the coarse structure of the UV candidate's K\"ahler potential might be, we give perturbative and exact examples which reproduce the kinetic functions under study in this work. We then turn our attention to String theory and argue that the necessary terms may be obtained. Specifically, we recall that the more general form of string loop corrections to the volume moduli K\"ahler potential in string compactifications spoil the K\"ahler potential's log-structure, and we hence expect them to break the shift symmetry at large field ranges. We conclude by discussing our results, and point out that our steepening corrections generically produce a moderate loss of CMB power at large angular scales for which we give an analytical estimate. Appendix A reviews the explicit form of the $f(R)$ theory dual to a broken shift symmetry at large fields.

\section{Shift Symmetry Primer}\label{Primer}

A scalar potential arising from a weakly broken shift symmetry is known to provide for controlled trans-Planckian field excursion during slow-roll inflation. The shift symmetry is said to be weakly broken, if the inflaton scalar potential $V(\phi)\ll 1$ itself provides the leading source of symmetry breaking. We begin our discussion by looking at the simple example of a scalar field minimally coupled to gravity
\beq\label{Sinf}
 \mathcal L= \sqrt{-g}\thinspace \left[ \tfrac12 R - \tfrac12 (\partial\phi)^2-V_0(\phi) \right] \thinspace.
\eeq
Generically, integrating out heavy fields and/or radiative contributions lead us to expect an infinite series of corrections to $V_0(\phi)$, containing in particular pieces of the form
\beq\label{Veff}
\Delta V= V_0(\phi)\,\sum\limits_{n\geq 1} c_n\frac{\phi^n}{\mpl^n}\thinspace.
\eeq
In the spirit of Wilsonian effective field theory (EFT) we should assume $c_n={\cal O}(1)\;\forall\;n$. Then we have
\begin{align}\label{etaproblem}
\left.\Delta \eta \right|_{\phi=\Delta\phi}&=  \sum\limits_{n\geq2}c_n n (n-1) \frac{\Delta\phi^{n-2}}{\mpl^{n-2}}\quad\gtrsim\quad 1
\end{align}
as soon as $\Delta\phi\gtrsim \mpl$. Controlling such trans-Planckian field excursions requires
\beq\label{shiftsymmCoeffs1}
c_n\thinspace\lesssim \thinspace\eta_0 \frac{1}{n^2} \frac{\mpl^{n-2}}{\Delta\phi^{n-2}}\thinspace\lesssim \thinspace 1
\eeq
which amounts to a shift symmetry $\phi\to\phi+c$ weakly broken by $V_0(\phi)$. Indeed, in the Lagrangian eq.~\eqref{Sinf}, the radiative corrections driven by the self-interactions (cubic and higher derivatives of $V_0$) fall off as $1/\phi^m,m>0$ at large $\phi>\mpl$, while graviton-loop corrections produce
\beq\label{VeffQG}
\delta V_{grav.}\sim V_0^m\,,\, (V_0'')^n \quad \ll \quad V_0
\eeq
provided $V_0\ll 1$.

The same shift symmetry requirement is evident in the structure of $f(R)$-gravity versions producing inflation models close to the Starobinsky model. Here, we need to require
\beq\label{feffR}
f(R)=R+\frac{c_2}{\mpl^2}R^2+\sum\limits_{n\geq3}\frac{c_n}{\mpl^{2n-2}}R^n
\eeq
with $c_2\gg 1$ and
\beq\label{shiftsymmCoeffs2}
\frac{c_n}{c_2^{n-1}}\ll \frac{1}{c_2^{n-1}}\ll 1\; \quad \forall\;n\geq 3\thinspace.
\eeq
Then the exponential approach to a shift-symmetric plateau potential
\beq\label{Vexp}
V(\phi)=V_0\left(1-e^{- \sqrt{2/3} \phi}+\ldots\right)
\eeq
for the associated canonically normalised scalar $\phi$ dominates the resulting scalar potential. The condition eq.~\eqref{shiftsymmCoeffs2} again marks the pattern of an effective weakly broken shift symmetry. For a more detailed discussion on shift symmetry and modified  gravity, see appendix \ref{a1}.

One of the main aims of this paper is to provide the analogue formulation of this shift symmetry for non-canonical models of inflation, to which we turn next.

\section{Pole Inflation}

\subsection{Laurent expansion}

Conventionally, inflation is studied in terms of a canonically normalised and minimally coupled scalar field and its effective potential. However, UV embeddings of inflationary physics often yield a minimally coupled but non-canonically normalised field, where the effective dynamics are encoded in the kinetic term as well as the field's potential. Instead of having inflationary dynamics only be determined by the scalar potential, a part of the models' complexity and predictivity now lies within the kinetic term. In particular, if the kinetic term may be cast as a Laurent series and given reasonable assumptions about the potential, one can study and understand inflationary dynamics solely in terms of the Laurent expansion's leading order pole and its residue~\cite{Galante:2014ifa}, as we will now recall. 

Consider an Einstein frame Lagrangian of the form
\begin{align} \label{KELagr}
 {\cal L}= \sqrt{-g}\thinspace \left[ \tfrac{1}{2} R-\tfrac{1}{2} (\frac{a_p}{\rho^p}+\ldots) \left(\partial\rho\right)^2-V_E\left(\rho\right) \right]
\end{align}
where we assume the kinetic function $K_E (\rho)$ to be given by a Laurent series with a pole of order $p$ at $\rho=0$ (without loss of generality) plus sub-leading terms, which are higher-order in $\rho$ (not higher order in $\rho^{-1}$) and are thus irrelevant close to the pole. In principle, higher order terms in $\rho^{-1}$, i.e.\ higher orders in the pole, are of increasing importance when $\rho\rightarrow \rho_0$. We will neglect those terms for now to ease our analysis of the first pole and give a condition which has to be satisfied in order to do so in \eqref{requirementK}.

The location of the pole corresponds to a fixed point of the inflationary trajectory, which is therefore characterised almost completely by this point. Upon canonical normalisation, the fixed point translates into a nearly shift-symmetric plateau in the potential. As the inflationary behaviour will be determined by the non-canonical field's trajectory in the vicinity of the pole, one may approximate $V_E(\rho)$ to be
\begin{equation} \label{V-Taylor}
V_E=V_0(1 - \rho+\ldots) \,,
\end{equation}
where the linear coefficient can be set to $-1$ without loss of generality. 
Sufficiently close to the pole at $\rho=0$, i.e.~at large $N$, the number of e-folds is related to the inflaton position as
 \begin{align}
  N  =   \frac{a_p}{(p-1)\rho^{p-1}}  \,, \quad 
  \rho = \left( \frac{a_p}{(p-1)N} \right)^{\frac{1}{p-1}} \,.
  \end{align}
Since we assume $p>1$, indeed the number of e-folds increases as one approaches the pole. Moreover, it is simple to calculate the slow-roll parameters,   
  \begin{align}
   \epsilon = \frac{1}{2 a_p} \rho^p \,, \quad
    \eta =  - \frac{p}{2 a_p} \rho^{p-1} \,.
   \end{align}
At lowest order in $1/N$, the inflationary predictions for this model are therefore given by
\begin{equation}
  n_s = 1 - \frac{p}{p-1} \frac1N \,, \qquad
  r=\frac{8 a_p^{\frac{1}{p-1}}}{(p-1)^{\frac{p}{p-1}}} \frac{1}{N^{\frac{p}{p-1}}} \,, \label{pole}
\end{equation}
where $N$ is the number of e-folds, $p$ the order of the pole in the kinetic function $K_E$ and $a_p$ is the leading coefficient of the Laurent expansion as in \eqref{KELagr}.

Putting all of this together, we observe that the presence of a fixed point of the kinetic function amounts to an \emph{effective shift symmetry} of the canonically normalised inflaton at large field values, provided that all higher-order poles in the kinetic function beyond the leading-order pole defining the fixed point have successively suppressed coefficients as in eq.~\eqref{requirementK} below. We emphasise that this happens regardless of the specific potential present in the non-canonical frame. This provides us with a new handle on finding regimes where inflaton potentials show an effective shift symmetry via analysing the local structure of the non-canonical kinetic function. In other words, vastly enhancing the kinetic term such that it becomes dominant with regard to the potential - e.g.\ with a pole in the kinetic function as above - enters the canonical normalisation such that, regardless of the specific potential, the canonically normalised field will slowly roll down its effective potential which will be of plateau type.

The case $p=2$ is special for a number of reasons. First of all, this gives rise to a value of the spectral index that agrees exceedingly well with the Planck data. Secondly, from a theoretical perspective, large classes of models with different interactions actually give rise to nearly identical predictions \eqref{pole} with $p=2$: this is referred to as the unity of cosmological attractors \cite{Galante:2014ifa}. In what follows, we will therefore reserve special emphasis for corrections to $p=2$ poles.

At this point we can finish the discussion of the effective shift symmetry arising near the pole in the kinetic function. We see that the single pole generates a scalar potential close to $\rho=0$ for the associated canonically normalised field $\phi$ of the form
\beq\label{Veff2}
V_0(\phi)=\left\{\begin{array}{c}V_0\,\left(1-A\,\phi^{\frac{2}{2-p}}\right)\quad,\quad p\neq 2 \\ \quad \\ V_0\,\left(1-\, e^{-\frac{\phi}{\sqrt{a_p}}}\right)\quad,\quad p=2\end{array}\right.
\eeq
where $A=\left(\frac{2-p}{2\sqrt{a_p}}\right)^{\frac{2}{2-p}}$. This shows the plateau at $\rho\to 0$ occurring for $\phi\to\infty$ if $p\geq2$ and for $\phi\to0$ otherwise. The higher powers in $\rho$ of $V_0(\rho)$ beyond the linear term are irrelevant due to the fact that the pole structure has inflation taking place for $\rho\to 0$. Hence, higher powers in the Laurent expansion of $K_E(\rho)$
\beq\label{KEeff1}
K_E(\rho)=\frac{a_p}{\rho^p}+\sum\limits_{q>p}\frac{a_q}{\rho^q}
\eeq
will perturb $V_0(\rho)\to V(\rho)=V_0(\rho)+\Delta V(\rho)$. Therefore, an extended plateau in the potential eq.~\eqref{Veff2} requires us to restrict to the regime where the following condition holds
\begin{equation}\label{requirementK}
\frac{a_q}{\rho^q}\ll \frac{a_p}{\rho^p}\,\,\,\forall\,\,\,q>p.
\end{equation}
Similar to the suppression of higher-dimension operators in some scalar potential, there is a priori no reason why condition \eqref{requirementK} should hold. We hence propose condition \eqref{requirementK} as a statement dual to the requirement to suppress higher-dimension operators in the canonical picture and will give a toy model realisation of \eqref{requirementK} in section \ref{complesPoles} and specifically via expression \eqref{perturbedkineticfunction}.
This suppression pattern of the residues of the Laurent expansion dictated by the approximate shift symmetry on the plateau forms a complete analogue of the two known cases discussed in the previous section.

\subsection{Universal corrections}\label{complexpoles}

Previously, we discussed how a real pole corresponding to a fixed point in field space translates to an approximately shift-symmetric plateau upon canonical normalisation. We now proceed to investigate the effect of perturbing the duality between fixed points and shift-symmetry, i.e.\ perturbing the function $K_E$. 

Suppose there is a higher-order pole with small coefficient $a_q$, (i.e. in the regime of validity of equation (\ref{requirementK})):
 \begin{align} \label{two-poles}
  K_E(\rho) = \frac{a_q}{\rho^q}+ \frac{a_p}{\rho^p} + \ldots \,,
 \end{align}
while the scalar potential is still given by the Taylor expansion, and the dots represent less singular terms in $\rho$. This gives rise to the relation
 \begin{align}
  N  =   \frac{a_q}{(q-1) \rho^{q-1}} + \frac{a_p}{(p-1)\rho^{p-1}} \quad
  \end{align}
in the approximation of being close to the pole. However, in order to invert this relation, one has to assume that the perturbation is small with respect to the original term:
 \begin{align}\label{condition38}
   \frac{a_q}{\rho^q} \ll \frac{a_p}{\rho^p} \,.
 \end{align}
We thus rediscover condition \eqref{requirementK}. As a perturbative expansion, we then obtain the solution:
 \begin{align}
   \rho = \rho_0 + \delta \rho \,, \quad \delta \rho = \frac{a_q}{a_p (q-1)} \rho_0^{p-q+1} \,,
 \end{align}
where $\rho_0$ (and subsequent quantities with the same subscript) refer to the unperturbed result of the previous section. Similarly, the corrections for the slow-roll parameters become
  \begin{align}
   \delta \epsilon & =    - \frac{(q-p-1)}{2 (q-1)} \frac{a_q}{a_p^2} \rho_0^{2p-q}  \,, \notag \\
    \delta \eta &  = - \frac{(q-p)(q-p-1)}{2 (q-1)} \frac{a_q}{a_p^2} \rho_0^{2p-q-1} \,,
    \end{align}
at lowest order in $a_q$. We therefore obtain
 \begin{align}
   \delta n_s & =  - \frac{a_q}{a_p{}^{\frac{q-1}{p-1}}} \frac{(q-p)(q-p-1)}{(q-1)(p-1)^{\frac{q-1}{p-1}-2}} N^{\frac{q-1}{p-1}-2} \,, \notag \\
  \delta r & = - \frac{8 a_q}{a_p{}^\frac{q-2}{q-1}} \frac{(q-p-1)}{(q-1)(p-1)^{\frac{2p-q}{p-1}}} N^{\frac{q-2p}{p-1}} \thinspace.
 \end{align}
These are universal corrections arising from a perturbation of the shift symmetry at large field values of the canonically normalised inflaton field.
 
Motivated both by observational and theoretical evidence (as well as the desire to obtain more manageable formulae), we now restrict ourselves to the case $p=2$. Other cases will be qualitatively identical. In the case of $p=2$ the above formulae simplify to
 \begin{align}
   n_s & = 1 - \frac{2}{N}  - \frac{a_q}{a_p{}^{q-1}} \frac{(q-2)(q-3)}{(q-1)} N^{q-3} \,, \notag \\
  r & = \frac{8 a_p}{N^2} - \frac{8 a_q}{a_p{}^\frac{q-2}{q-1}} \frac{(q-3)}{(q-1)} N^{q-4} \,.
  \label{pert-pred}
 \end{align}
These are the universal corrections to the cosmological attractor predictions provided the corrections respect the effective shift symmetry structure of eq.~\eqref{requirementK}. They should be understood as a double expansion, both in $1/N$ as well as in $a_q N^{q-2}$. The latter requirement follows from the approximation to obtain $\rho(N)$ (and is given by $a_q N^{\frac{q-p}{p-1}}$ in general). Here we have assumed that $a_p$ is of order one. In terms of these expansion parameters, the correction term to the spectral index is bilinear in both, while the correction to the tensor-to-scalar ratio is an order in $1/N$ higher. Corrections bilinear in $N$ will become of increasing importance for larger $N$. This will nicely be illustrated in the next subsection when transforming to canonical fields.

\subsection{Canonical Formulation}

We will now turn to a description of the corrections to the plateau potential arising from the least suppressed residue $a_q$ in the Laurent expansion. Starting from the perturbed Laurent expansion with two poles \eqref{two-poles},  we find the relation $\rho(\phi)$ for the canonically normalised field $\phi$ for $p\neq 2$ to leading order in $a_q$ as
\beq\label{CanNonCan1}
\rho(\phi)= A\, \phi^{\frac{2}{2-p}} + \frac{a_q}{2\sqrt{a_p}} A^{\frac{p-2(q-2)}{2}} \phi^{\frac{2}{2-p} (p-q+1)}\quad.
\eeq
For the special case $p=2$ the relation becomes exponential and we get
\beq
\rho(\phi)=e^{-\frac{\phi}{\sqrt{a_p}}} + \tfrac{1}{4} a_q e^{(q-3)\frac{\phi}{\sqrt{a_p}}}\quad,\quad q > p =2\quad.
\eeq
We obtain the resulting canonical scalar potential $V(\phi)=V_0(\rho(\phi))$ to ${\cal O}(a_q)$ by plugging eq.~\eqref{V-Taylor} into the original potential 
\beq\label{Veff3}
V(\phi)=\left\{\begin{array}{c}V_0\,\left(1- A\,\phi^{\frac{2}{2-p}} - a_q \,B\, \phi^{\frac{2}{2-p} (p-q+1)}\right)\;,\; p\neq 2  \\ \\ V_0\,\left(1- e^{-\frac{\phi}{\sqrt{a_p}}} - \tfrac14 a_qe^{(q-3)\frac{\phi}{\sqrt{a_p}}}\right)\;,\; p=2\quad.\end{array}\right.
\eeq
Here, the correction coefficient $B$ for $p\neq 2$ has the form $B=\frac{1}{2\sqrt{a_p}} A^{\frac{p-2(q-2)}{2}}$. 
Consequently, for $a_q<0$ the plateau potential universally acquires a rising correction, increasing for $\phi\to\infty$ for $p\geq2$, and for $p<2$ rising towards $\phi=0$.

We see that if all microscopic parameters $a_p,p,q$ take ${\cal O}(1)$ values, then $A,B={\cal O}(1)$ as well, and their precise values are irrelevant for the general arguments given here. The only relevant quantities are:
\begin{itemize}
\item $p$ which determines the leading functional form of the plateau potential, 
\item $a_q\ll 1$ which controls the magnitude of the correction, and 
\item  the difference $q-p$ which controls the functional dependence of the first correction in the scalar potential.
\end{itemize}
This structure of the scalar potential allows for two observations. 

First of all, the case
\beq
q=p+1
\eeq
leads to a rather curious observation. Namely, in this case the corrections to the  scalar potential are constant! For this reason, they only serve to redefine the constants in the original form \eqref{Veff2}:
\begin{align} \label{Veff5}
& \tilde A=\frac{A}{1 - a_q B}\;,\; \tilde V_0=V_0\,(1 - a_q B)\quad,\quad p\neq 2  \\
& \tilde V_0=V_0\,(1 - \tfrac{1}{4} a_q)\quad,\quad p=2 \,.
\end{align}
There is an easy way to understand why this happens. We look again at the kinetic function, writing 
\beq\label{KEeff1b}
K_E(\rho)=\frac{a_p}{\rho^p}+\frac{a_{p+1}}{\rho^{p+1}}+\sum\limits_{q>p+1}\frac{a_q}{\rho^q}\quad.
\eeq
Now perform a field redefinition $\rho\to\rho+\varepsilon$, and insert this into $K_E$. We get
\begin{align}\label{KEeff1c}
K_E(\rho+\varepsilon)&=\frac{a_p}{(\rho+\varepsilon)^p}+\frac{a_{p+1}}{(\rho+\varepsilon)^{p+1}}+\ldots\nonumber\\
&= \frac{a_p}{\rho^p}-\frac{p a_p\varepsilon}{\rho^{p+1}}+\frac{a_{p+1}}{\rho^{p+1}}+\ldots
\end{align}
Hence, by adjusting the field definition to $\varepsilon=a_{p+1}/(pa_p)$ we can \emph{always} absorb the pole of order $p+1$, but not other poles of higher-order at the same time. This is the reason why a pole at order $p+1$ does not contribute at that order to the scalar potential. Beyond leading order, the field redefinition generates contributions to poles at order $p+2$ and higher. Consequently, we expect a pole in $K_E$ at order $p+1$ to contribute at higher sub-leading orders to the scalar potential. 

At the next order, i.e.~for
\beq
q=p+2
\eeq
the correction scales as the inverse of the leading plateau potential term. The cases $p=1,2,3$ form illustrative examples for this situation
\beq\label{Veff4}
V(\phi)=\left\{\begin{array}{c}V_0\,\left(1- A\,\phi^2 - a_q \,\frac{B}{\phi^2}\right)\quad,\quad p=1  \\0 \\ V_0\,\left(1- e^{-\frac{\phi}{\sqrt{a_p}}} - \tfrac14 a_qe^{\frac{\phi}{\sqrt{a_p}}}\right)\quad,\quad p=2 \\0 \\ V_0\,\left(1-\frac{A}{\phi^2} - a_q \,B\,\phi^2\right)\quad,\quad p=3\quad. \end{array}\right.
\eeq

\section{Complex Poles}\label{complesPoles}

\begin{figure}
    \centering
    \includegraphics[scale=0.5]{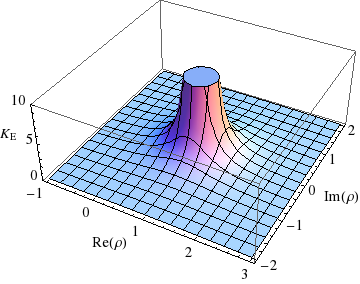}\quad\quad\quad \includegraphics[scale=0.5]{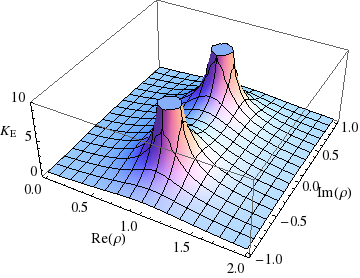}
    \caption{\emph{Pole structure of $K_E$.} \textbf{Left panel:} \emph{a pole of order $p=2$, localised in the real part of $\rho$.} \textbf{Right panel:} \emph{the perturbed case, showing the split of the original pole in two complex poles of order one. The non-canonical inflaton may now move over the hilltop along the real line which corresponds to shift-symmetry breaking in canonical fields.}}
    \label{complexpolestructure2}
\end{figure}

The above analysis only considers a single correction. In order to further link the above discussion with our argument in chapter \ref{Primer}, we will now consider an infinite tower of corrections to the leading pole of a kinetic function and hence readily demonstrate that in order not to spoil the inflationary dynamics, a hierarchy between the corrections reminiscent of the EFT argument has to arise. To that end, consider higher powers in the Laurent expansion of $K_E(\rho)$
\beq\label{KEeff1}
K_E(\rho)=\frac{a_p}{\rho_p}+\sum\limits_{q>p}\frac{a_q}{\rho^q}\quad .
\eeq
As an example, we will assume the above to arise from a toy model of the closed form
\begin{equation}\label{perturbedkineticfunction}
K_E(\rho)=\frac{a_p}{\rho^2+\varepsilon^2}\quad,
\end{equation}
where $\varepsilon^2\ll 1$. The perturbation $\varepsilon^2$ affects the pole structure, moving the pole at $\rho_0=0$ from the real to the complex plane at $\rho_0\rightarrow \pm i\varepsilon$, as shown in figure \ref{complexpolestructure2}. It is important to note that the function $K_E(\rho)$ does not become complex itself at any point, it merely contains a complex pole. 

The inflationary predictions from the presence of a complex pole follow at lowest order from the universal corrections that we derived earlier: expanding the complex pole
  \begin{align}\label{expansion}
  K_E = \frac{a_p}{\rho^2} - \frac{a_p \varepsilon^2}{\rho^4} + \frac{a_p \varepsilon^4}{\rho^6} + \ldots \,,
      \end{align}
it is clear that at lowest order in $\varepsilon^2$ the form of the kinetic function, and hence the inflationary predictions, is exactly that of the perturbed Laurent expansion considered previously with $p=2$ and $q=4$. As a consequence, the inflationary predictions are given by \eqref{pert-pred} with $q=4$ and $a_q = - a_p\varepsilon^2$. Note that the latter always corresponds to a blue-shifting of the spectral index at large $N$. Further note how \eqref{expansion} realises \eqref{KEeff1} with $a_q\ll a_p\,\forall\,q>p$, i.e.\ satisfies \eqref{requirementK}. This suppression pattern of the residues of the Laurent expansion dictated by the approximate shift symmetry on the plateau is in complete analogy to the known earlier cases given in chapter \ref{Primer}.

If the above kinetic term only has a complex pole (and no sub-leading corrections), the transition to a canonical inflaton field $\phi$ can be done exactly and reads
\begin{equation}\label{exactsolution}
\rho = e^{-\phi/\sqrt{a_p}}- \tfrac14 \varepsilon^2 e^{\phi/\sqrt{a_p}} .
\end{equation}
The scalar potential around the would-be pole reads
\begin{equation}
V_E=V_0( 1 - e^{-\phi/\sqrt{a_p}} + \tfrac14 \varepsilon^2 e^{\phi/\sqrt{a_p}} +\ldots) \,.
 \label{risingStaro}
\end{equation}
Again, the nearly shift-symmetric plateau of the canonically normalised inflaton is broken at large field values, the exact value depending on the perturbation $\varepsilon^2$ of the kinetic pole structure. 

\section{Towards a UV embedding}

\subsection{K\"ahler potentials}

Within the framework of non-canonical inflation, the complexity of the inflationary dynamics has been shifted to the kinetic function. Poles in the kinetic function then translate to nearly shift-symmetric potentials and complex poles or higher order terms in $1/\rho$ break the shift-symmetry at large fields. It is therefore an important question whether one can embed kinetic functions with the aforementioned structure into a UV theory. We start with a general observation.

Consider a toy potential of the form $K = \log f$
where the function $f$ in the argument of the logarithm is a real function, e.g.\ an arbitrary polynomial, of  $\Phi+\bar\Phi$ or $\Phi\bar\Phi$. Now assume that $f$ has a real zero of order $n$ at e.g.~$\Phi_0 = 0$. Close to the pole, the function can then be approximated as  $f= (\Phi + \bar \Phi)^n + \ldots$. The corresponding K\"ahler metric takes the following form
\begin{equation}\label{41}
K_{\Phi\bar{\Phi}} = \frac{f_\Phi f_{\bar \Phi} - f_{\Phi \bar \Phi} f}{f^2} =  \frac{n}{( \Phi + \bar \Phi  )^2} + \ldots \,.
\end{equation}
Upon identifying $\Phi = \bar \Phi = \rho$ and $n = 2 a_p$ this becomes the previously considered Laurent expansion. The order of the pole is therefore independent of the order of the zero in the argument of the logarithm; instead, the order of the zero determines the residue of the pole, which always has order two. This argument underlines the robustness of cosmological attractors: changes to the location of the zero of $f$ and to its order do not affect the resulting pole structure of order two. Given that the function $f$ has a zero of some order, the denominator of \eqref{41} always factorises on the real line.

Turning to the type of corrections corresponding to complex poles
 \begin{align}
  K_{\Phi\bar{\Phi}} &=\frac{n}{( \Phi + \bar \Phi  )^2 + \varepsilon^2}\nonumber\\
  & = \frac{n}{( \Phi + \bar \Phi  )^2} - \frac{n \varepsilon^2}{ ( \Phi + \bar \Phi  )^4} + \ldots \,,
  \label{pert-Kahler}
  \end{align}
that become relevant at large field values, we  first note that the denominator does not factorise on the real but only on the complex plane. Hence in order to generate such K\"ahler potentials, we must resort to a different structure than the one described above. A prototypical example would  be $f = ( \Phi + \bar \Phi  )^2 + \varepsilon^2$. Indeed this will induce additional terms in the K\"ahler metric that correspond to higher-order poles, similar to \eqref{pert-Kahler}:
 \begin{align}
  K_{\Phi\bar{\Phi}} &=\frac{n (( \Phi + \bar \Phi  )^2 - \varepsilon^2) }{(( \Phi + \bar \Phi  )^2 + \varepsilon^2)^2} \nonumber\\
  &= \frac{n}{( \Phi + \bar \Phi  )^2} - \frac{3 n \varepsilon^2}{ ( \Phi + \bar \Phi  )^4} + \ldots \,.
  \label{pert-Kahler2}
 \end{align}
In order to obtain exactly the  K\"ahler metric \eqref{pert-Kahler2}, we note that it can actually be integrated to yield
\begin{equation}\label{45}
K=\frac{\Phi+\bar\Phi}{\varepsilon}\tan ^{-1}\left(\frac{\Phi+\bar\Phi }{\varepsilon }\right)-\frac{1}{2} \ln \left(\varepsilon ^2\hspace{-0.3ex}+\hspace{-0.3ex}(\Phi+\bar\Phi)^2\right).
\end{equation}
Expanding the K\"ahler potential at small $\varepsilon$, we find
 \begin{align}
  K = -n \log( \Phi + \bar \Phi) - \frac{n \varepsilon^2}{6 (\Phi + \bar \Phi)^2} + \ldots \,.
 \end{align}
The leading term outside the logarithm corresponds to the pole of order four (necessarily with opposite sign, to counter the pole of order two) that is the first to become relevant at large field values, i.e.~at large $N$. As we have argued, this gives rise to a universal signature in terms of the spectral index and tensor-to-scalar ratio.

An alternative way to generate the type of corrections that become relevant at large field values is to simply consider corrections to the K\"ahler potential outside of the logarithm, i.e.~a more general expansion of the form \eqref{exp-pole} but with arbitrary corrections and order:
 \begin{align}
  K = -n \log( \Phi + \bar \Phi) + \frac{n'}{(\Phi + \bar \Phi)^{q-2}} + \ldots \,. \label{exp-pole}
 \end{align}
In a way, terms outside of the logarithm may be thought of as a way to reintroduce the $\eta$-problem, but in a controlled way such that this only occurs at large fields. As we have argued, this gives rise to a universal signature in terms of the spectral index and tensor-to-scalar ratio at leading order in $a_q = 2^{-q} (q-1)(q-2) n'$.
 
\subsection{Comments on matching to string theory}
 
Which of these structures can be obtained in string theory settings? If we look at the peculiar behaviour of non-canonical inflation with $p=2$ and $q=3$, we discover by comparison a relation to a well known fact of the K\"ahler geometry argued for 1-loop corrections in string theory to the volume moduli K\"ahler potential $K$ in supergravity~\cite{vonGersdorff:2005bf,Berg:2005ja,Berg:2007wt,Cicoli:2007xp,Berg:2014ama}. Namely, for $p=2$ we can think of $K_E(\rho)$ as arising from a logarithmic K\"ahler potential for a chiral modulus field $\chi$
\beq
K_0=-2a_p \ln(\chi+\bar\chi)\quad,\quad \chi+\bar\chi=2\rho
\eeq
where we get $K_E(\rho)=\partial_\chi\partial_{\bar\chi}K\equiv K_{\bar\chi\chi}$. A string loop correction to $K$ is the generically argued~\cite{vonGersdorff:2005bf,Berg:2005ja,Berg:2007wt,Cicoli:2007xp,Berg:2014ama} to change $K$ with a quantity
\begin{align}
 \delta K = -\frac{2^q}{(q-2)(q-1)}\frac{a_q}{(\chi+\bar\chi)^{q-2}}\quad,\quad q=3,4\quad.
\end{align}
Here, we have chosen the prefactor of the loop correction such that the induced term in $K_{\bar\chi\chi}$ matches the form eq~\eqref{two-poles}. Hence, according to~\cite{vonGersdorff:2005bf,Berg:2005ja,Berg:2007wt,Cicoli:2007xp,Berg:2014ama} the corrections form degree $-(q-2)$ polynomials in $K$. From the general analysis in~\cite{Berg:2007wt,Cicoli:2007xp} we know that for constant superpotential $W_0$ the leading-order supergravity scalar potential for such a modulus $\chi$ induced by the above K\"ahler potential correction scales like
\beq
\delta V \sim (2-q) (3-q) \delta K\quad.
\eeq
Again, we see that for $q=3$ the leading correction to the potential vanishes. In this context of string loop corrections in type IIB compactifications this phenomenon was named ``extended no-scale structure'' in~\cite{vonGersdorff:2005bf,Berg:2005ja,Berg:2007wt,Cicoli:2007xp} as the above leading correction to the no-scale potential of the K\"ahler moduli (which have $a_{p=2}=3/2$) was observed there to vanish (and hence ``extend'' no-scale) for all loop corrections to $K$ which scale with power $q=3=p+1$ in the resulting K\"ahler metric $K_{\bar\chi\chi}$.

Our analysis of the scalar potential above shows that for models with pole-dominated kinetic terms this extended no-scale structure holds for kinetic functions with an arbitrary leading pole of order $p>0$ even if $p\neq 2$. Moreover, it has a natural explanation as a shift redefinition of the modulus.

We can now take a look a the leading-order structure of both the string 1-loop and the leading ${\cal O}(\alpha'^3)$-corrections to the type IIB volume moduli K\"ahler potential
\begin{align}
K&=-2\ln({\cal V}+\xi/2)-\frac{C}{T+\bar T}-\frac{D}{(T+\bar T)^2} \,, \nonumber \\
&=-2\ln {\cal V}-\frac{\xi}{(T+\bar T)^{3/2}}-\frac{C}{T+\bar T}-\frac{D}{(T+\bar T)^2} \,,
\end{align}
with ${\cal V}\sim (T+\bar T)^{3/2}$ and at lowest order in $\xi$. In such a simple situation of a single K\"ahler modulus the above inflationary regime would correspond to working close to $T=0$ where the $\alpha'$-corrections are out of control. However, the simple toy example serves us here to point out that a comparison with string theory as a possible UV completion fixes concrete numbers for the possible values for $p$ and $q$. Namely, from the single modulus toy example we get $p=2$ and $q=3,7/2,4$ of which the $q=3$ contribution drops out of the scalar potential at leading order as discussed above. Moreover, matching to a string example would allow us also to compute the compute $c$ and $a_q$ in terms of the microscopic parameters $\xi, C,D$. As $C,D$ are $g_s$-suppressed in the string coupling compared to the tree level terms and $\xi$, this may allow also for an understanding of the smallness of $a_q$ in terms of small $g_s$. It remains to be seen, whether an embedding of this structure in a concrete controlled string theory setting (either away from small volume regimes, or in a better-controlled singular regime) is possible.

\section{Discussion}\label{concl}

The topic of this paper is non-canonical inflation. We have recalled how a leading pole in the Laurent expansion of the kinetic function translates into a nearly shift-symmetric plateau in the effective scalar potential of the canonically normalised inflaton field, i.e.~a fixed point of the cosmological evolution. This is a generic feature and does not depend on the order of the pole. 

Subsequently, we have investigated higher-order poles as perturbations of the Laurent expansion of the kinetic term. The fixed point hence vanishes which results in the approximate shift-symmetry of the inflaton potential to be broken at large fields. Given a hierarchical suppression of higher order poles, we have outlined the leading corrections to the inflationary predictions in terms of the number of e-folds and the perturbation of the pole structure, and found that such corrections induce terms with positive powers of $N$ in the spectral index $n_s$, which therefore rise to prominence at sufficiently large-$N$ (i.e.~at large field values). Moreover, we have provided an explanation of the irrelevance of the first higher-order pole and have argued this to be an alternative way to view the extended no-scale structure in string theory: the effect of a pole one order higher than the leading one can be absorbed in a redefinition of the field.

We can use our results for $n_s$ to analytically estimate the power-loss at large angular scales resulting from the blue-shifting of the spectral index. From the definition of $n_s=d \ln P/ d\ln k =P^{-1} d P/ d N$ we get
\begin{align}
  \left.\frac{\Delta P(\delta n_s)}{P}\right|^{N}_{N+\Delta N}=\int\limits_{N+\Delta N}^{N} \delta n_s &\notag\\ &\notag \\
    =   \frac{a_q}{a_p{}^{\frac{q-1}{p-1}}}   \frac{(q-p)(q-p-1)}{(q-1)(p-1)^{\frac{q-2p+1}{p-1}}} & N^{\frac{q-2p+1}{p-1}} \Delta N+{\cal O}(\Delta N^2)\;.
 \end{align}
For the particular case of exponential potentials arising from $p=2$ and $q=4$, setting $a_2=1$ and $a_4=-\varepsilon^2$ we get
\begin{align}
  \left.\frac{\Delta P(\delta n_s)}{P}\right|^{N}_{N+\Delta N}=&-\frac23 \varepsilon^2  N \Delta N+{\cal O}(\Delta N^2)\;.
 \end{align}
In the same particular case we also have that
\beq
\delta n_s=\frac23 \varepsilon^2 N
\eeq
Using $N=60$ we see that a bound $\varepsilon^2\lesssim 2\times 10^{-4}$ limits the shift of the spectral index to $\delta n_s \lesssim 0.008$ which is the 2-$\sigma$ range for the $n_s$ measurement from Planck. By plugging in these numbers and the range of efolds $\Delta N\simeq 5$ over which power-loss occurs we find the power loss for this case to be
\beq
\left.\frac{\Delta P(\delta n_s)}{P}\right|^{N}_{N+\Delta N}=-\frac23 \varepsilon^2  N \Delta N\simeq - 0.04
\eeq
which is about $4$\%.  This is in qualitative agreement with previous studies employing exponentially rising corrections \cite{0303636, 1211.1707, Cicoli:2013oba, 1309.3413, 1309.4060, 1404.2278, Kallosh:2014xwa, Cicoli:2014bja, PhysRevD.91.023514}. For order-one values of $a_p$, $p$ and $q$, one obtains similar results. 

Finally, we have discussed the possible UV embedding
of non-canonical inflation. While K\"ahler potentials of
logarithm type are bread and butter in string theory compactifications,
loop corrections can induce higher-order
terms in the K\"ahler potential. These would generically
result in a shift symmetry breaking at large field displacements.
We leave a concrete embedding of these terms into
a reliable string theory set-up for future investigation. In particular, the properties of complex structure moduli space close to a conifold point may provide a viable path to embedding our structure into string theory, while working with volume moduli close to zero volume (if one took the above toy comparison literally) is clearly a badly controlled regime.

While elegant, rephrasing the formulation of inflationary dynamics in terms of a non-trivial kinetic function does not reduce the severity of the need to fine tune the scenario. Having to impose condition \eqref{requirementK} is one of our central findings. Further, one formulation may not be understood as more fundamental than the other. However, it recasts the context in which the fine tuning has to occur such that new insights may be possible. Concretely, since a non-canonical kinetic term arises an intermediate step in any UV derived 4D effective theory, it is instructive and may even provide a short cut to know which types of non-canonical kinetic terms affect possible inflationary dynamics in what way.

\acknowledgments

BB and AW are supported by the Impuls und Vernetzungsfond of the Helmholtz Association of German Research Centres under grant HZ-NG-603. BB appreciates the RUG HEP theory group's hospitality during his stay while this work was in preparation. MG is grateful to the DESY theory group, where a part of this work was done. Finally, DR and AW acknowledge the support in part by National Science Foundation Grant No.~PHYS-1066293 and the stimulating hospitality of the Aspen Center for Physics, as well as useful discussions with Cliff Burgess.

\appendix

\section{Shift symmetry and $f(R)$}\label{a1}

It is well known that potentials such as 
\begin{equation}
V\sim V_0\left(1-e^{-\sqrt{2/3}\chi}+\ldots \right)
\end{equation}
may be recast in terms of an $f(R)\sim R^2$ - theory in the inflationary regime given sub-leading terms die off exponentially fast such that $V\rightarrow V_0 = constant$ for large fields. 

The study of single higher order terms such as e.g.\ $\Delta f(R)\sim \beta R^3$ has shown either to induce a runaway direction towards large fields in the potential ($\beta>0$) or to cause instabilities that lead to causally disconnected spacetime regions ($\beta<0$) \cite{BandM}. It is crucial to note that regardless of the suppression of the single higher order term, i.e.\ the size of the coefficient $\beta$, the resulting effect will eventually occur, thus rendering the resulting scalar field potential ultimately unstable or ill-defined (depending on the sign of $\beta$ respectively). 

Now consider the vanilla Starobinsky potential where a rising exponential has been added
\begin{equation}
V=V_0\left(1-e^{-\sqrt{2/3}\phi}\right)^2+\varepsilon V_0 e^{\sqrt{2/3}\phi} -\varepsilon\thinspace V_0 \thinspace.
\end{equation}
This toy model displays a nearly shift-symmetric plateau in an intermediate regime which is broken at large fields given $\varepsilon\ll 1$ (the size of $\varepsilon$ determines the length of the inflationary plateau). The subtraction of the term $\varepsilon\thinspace V_0$ ensures a minimum at $\phi=0$.

It can be shown \cite{Broy:2014xwa}, that the above symmetry breaking at large fields may exactly be recast in the language of modified gravity as
\begin{equation}\label{risingF}
f(R)=\frac{\varepsilon-1}{3\varepsilon}R+4\varepsilon V_0\left[\frac{(1-\varepsilon)^2}{9\varepsilon^2}+\frac{2}{3\varepsilon}+\frac{R}{6\varepsilon V_0} \right]^{3/2}
\end{equation}
plus some integration constant which may readily be obtained. Expanding the above then precisely recovers the pattern described by \eqref{feffR} with an \emph{infinite} tower of higher order terms suppressed with appropriate powers of $\varepsilon$. The coefficients of the first and second term are easily shown to approach the familiar Starobinsky coefficients for $\varepsilon\rightarrow 0$ while all the higher order terms vanish in that limit. 

We thus learn that an enhanced $c_2 R^2$ term in the series expansion of a closed form that is to leading order $R^n$ with $n<2$ is responsible for the inflationary plateau, as expected, where the shift symmetry breaking is realised by an infinite but suppressed tower of higher order terms. It is important to note, that there has to be an infinite number of higher order terms to avoid the drastic consequences described to come from a single higher order term or, likewise, a finite series.

\bibliography{alpha}

\providecommand{\href}[2]{#2}\begingroup\raggedright\begin{thebibliography}{10}

\bibitem{1212.5225}
{\bf WMAP} Collaboration, C.~Bennett et~al., {\it {Nine-Year Wilkinson
  Microwave Anisotropy Probe (WMAP) Observations: Final Maps and Results}},
  {\em Astrophys.J.Suppl.} {\bf 208} (2013) 20,
  [\href{http://arxiv.org/abs/1212.5225}{{\tt arXiv:1212.5225}}].

\bibitem{Ade:2015lrj}
{\bf Planck} Collaboration, P.~Ade et~al., {\it {Planck 2015 results. XX.
  Constraints on inflation}},  \href{http://arxiv.org/abs/1502.02114}{{\tt
  arXiv:1502.02114}}.

\bibitem{Ade:2015tva}
{\bf BICEP2, Planck} Collaboration, P.~Ade et~al., {\it {Joint Analysis of
  BICEP2/$Keck  Array$ and $Planck$ Data}},  {\em Phys.Rev.Lett.} {\bf 114}
  (2015), no.~10 101301, [\href{http://arxiv.org/abs/1502.00612}{{\tt
  arXiv:1502.00612}}].

\bibitem{Guth:1980zm}
A.~H. Guth, {\it {The Inflationary Universe: A Possible Solution to the Horizon
  and Flatness Problems}},  {\em Phys.Rev.} {\bf D23} (1981) 347--356.

\bibitem{Linde:1981mu}
A.~D. Linde, {\it {A New Inflationary Universe Scenario: A Possible Solution of
  the Horizon, Flatness, Homogeneity, Isotropy and Primordial Monopole
  Problems}},  {\em Phys.Lett.} {\bf B108} (1982) 389--393.

\bibitem{Albrecht:1982wi}
A.~Albrecht and P.~J. Steinhardt, {\it {Cosmology for Grand Unified Theories
  with Radiatively Induced Symmetry Breaking}},  {\em Phys.Rev.Lett.} {\bf 48}
  (1982) 1220--1223.

\bibitem{Linde:1983gd}
A.~D. Linde, {\it {Chaotic Inflation}},  {\em Phys.Lett.} {\bf B129} (1983)
  177--181.

\bibitem{Kawasaki:2000yn}
M.~Kawasaki, M.~Yamaguchi, and T.~Yanagida, {\it {Natural chaotic inflation in
  supergravity}},  {\em Phys. Rev. Lett.} {\bf 85} (2000) 3572--3575,
  [\href{http://arxiv.org/abs/hep-ph/0004243}{{\tt hep-ph/0004243}}].

\bibitem{Lyth:1996im}
D.~H. Lyth, {\it {What would we learn by detecting a gravitational wave signal
  in the cosmic microwave background anisotropy?}},  {\em Phys.Rev.Lett.} {\bf
  78} (1997) 1861--1863, [\href{http://arxiv.org/abs/hep-ph/9606387}{{\tt
  hep-ph/9606387}}].

\bibitem{Garcia-Bellido:2014wfa}
J.~Garcia-Bellido, D.~Roest, M.~Scalisi, and I.~Zavala, {\it {Lyth bound of
  inflation with a tilt}},  {\em Phys.Rev.} {\bf D90} (2014), no.~12 123539,
  [\href{http://arxiv.org/abs/1408.6839}{{\tt arXiv:1408.6839}}].

\bibitem{Starobinsky}
A.~Starobinsky, {\it {A New Type of Isotropic Cosmological Models Without
  Singularity}},  {\em Physical Letters B} {\bf 91} (1980), no.~1 99.

\bibitem{conformal-inflation}
R.~Kallosh and A.~Linde, {\it {Universality Class in Conformal Inflation}},
  {\em JCAP} {\bf 1307} (2013) 002, [\href{http://arxiv.org/abs/1306.5220}{{\tt
  arXiv:1306.5220}}].

\bibitem{Ferrara:2013}
S.~Ferrara, R.~Kallosh, A.~Linde, and M.~Porrati, {\it {Minimal Supergravity
  Models of Inflation}},  {\em Phys.Rev.} {\bf D88} (2013), no.~8 085038,
  [\href{http://arxiv.org/abs/1307.7696}{{\tt arXiv:1307.7696}}].

\bibitem{Kallosh:2013tua}
R.~Kallosh, A.~Linde, and D.~Roest, {\it {Universal Attractor for Inflation at
  Strong Coupling}},  {\em Phys.Rev.Lett.} {\bf 112} (2014), no.~1 011303,
  [\href{http://arxiv.org/abs/1310.3950}{{\tt arXiv:1310.3950}}].

\bibitem{Kallosh:2013yoa}
R.~Kallosh, A.~Linde, and D.~Roest, {\it {Superconformal Inflationary
  $\alpha$-Attractors}},  {\em JHEP} {\bf 1311} (2013) 198,
  [\href{http://arxiv.org/abs/1311.0472}{{\tt arXiv:1311.0472}}].

\bibitem{Galante:2014ifa}
M.~Galante, R.~Kallosh, A.~Linde, and D.~Roest, {\it {Unity of Cosmological
  Inflation Attractors}},  {\em Phys.Rev.Lett.} {\bf 114} (2015), no.~14
  141302, [\href{http://arxiv.org/abs/1412.3797}{{\tt arXiv:1412.3797}}].

\bibitem{vonGersdorff:2005bf}
G.~von Gersdorff and A.~Hebecker, {\it {Kahler corrections for the volume
  modulus of flux compactifications}},  {\em Phys.Lett.} {\bf B624} (2005)
  270--274, [\href{http://arxiv.org/abs/hep-th/0507131}{{\tt hep-th/0507131}}].

\bibitem{Berg:2005ja}
M.~Berg, M.~Haack, and B.~Kors, {\it {String loop corrections to Kahler
  potentials in orientifolds}},  {\em JHEP} {\bf 0511} (2005) 030,
  [\href{http://arxiv.org/abs/hep-th/0508043}{{\tt hep-th/0508043}}].

\bibitem{Berg:2007wt}
M.~Berg, M.~Haack, and E.~Pajer, {\it {Jumping Through Loops: On Soft Terms
  from Large Volume Compactifications}},  {\em JHEP} {\bf 0709} (2007) 031,
  [\href{http://arxiv.org/abs/0704.0737}{{\tt arXiv:0704.0737}}].

\bibitem{Cicoli:2007xp}
M.~Cicoli, J.~P. Conlon, and F.~Quevedo, {\it {Systematics of String Loop
  Corrections in Type IIB Calabi-Yau Flux Compactifications}},  {\em JHEP} {\bf
  0801} (2008) 052, [\href{http://arxiv.org/abs/0708.1873}{{\tt
  arXiv:0708.1873}}].

\bibitem{Berg:2014ama}
M.~Berg, M.~Haack, J.~U. Kang, and S.~Sjörs, {\it {Towards the one-loop
  K\"ahler metric of Calabi-Yau orientifolds}},  {\em JHEP} {\bf 1412} (2014)
  077, [\href{http://arxiv.org/abs/1407.0027}{{\tt arXiv:1407.0027}}].

\bibitem{0303636}
C.~R. Contaldi, M.~Peloso, L.~Kofman, and A.~D. Linde, {\it {Suppressing the
  lower multipoles in the CMB anisotropies}},  {\em JCAP} {\bf 0307} (2003)
  002, [\href{http://arxiv.org/abs/astro-ph/0303636}{{\tt astro-ph/0303636}}].

\bibitem{1211.1707}
S.~Downes and B.~Dutta, {\it {Inflection Points and the Power Spectrum}},  {\em
  Phys.Rev.} {\bf D87} (2013), no.~8 083518,
  [\href{http://arxiv.org/abs/1211.1707}{{\tt arXiv:1211.1707}}].

\bibitem{Cicoli:2013oba}
M.~Cicoli, S.~Downes, and B.~Dutta, {\it {Power Suppression at Large Scales in
  String Inflation}},  {\em JCAP} {\bf 1312} (2013) 007,
  [\href{http://arxiv.org/abs/1309.3412}{{\tt arXiv:1309.3412}}].

\bibitem{1309.3413}
F.~G. Pedro and A.~Westphal, {\it {Low-$\ell$ CMB power loss in string
  inflation}},  {\em JHEP} {\bf 1404} (2014) 034,
  [\href{http://arxiv.org/abs/1309.3413}{{\tt arXiv:1309.3413}}].

\bibitem{1309.4060}
R.~Bousso, D.~Harlow, and L.~Senatore, {\it {Inflation after False Vacuum
  Decay: Observational Prospects after Planck}},  {\em Phys.Rev.} {\bf D91}
  (2015), no.~8 083527, [\href{http://arxiv.org/abs/1309.4060}{{\tt
  arXiv:1309.4060}}].

\bibitem{1404.2278}
R.~Bousso, D.~Harlow, and L.~Senatore, {\it {Inflation After False Vacuum
  Decay: New Evidence from BICEP2}},  {\em JCAP} {\bf 1412} (2014), no.~12 019,
  [\href{http://arxiv.org/abs/1404.2278}{{\tt arXiv:1404.2278}}].

\bibitem{Kallosh:2014xwa}
R.~Kallosh, A.~Linde, and A.~Westphal, {\it {Chaotic Inflation in Supergravity
  after Planck and BICEP2}},  {\em Phys.Rev.} {\bf D90} (2014) 023534,
  [\href{http://arxiv.org/abs/1405.0270}{{\tt arXiv:1405.0270}}].

\bibitem{Cicoli:2014bja}
M.~Cicoli, S.~Downes, B.~Dutta, F.~G. Pedro, and A.~Westphal, {\it {Just enough
  inflation: power spectrum modifications at large scales}},  {\em JCAP} {\bf
  1412} (2014), no.~12 030, [\href{http://arxiv.org/abs/1407.1048}{{\tt
  arXiv:1407.1048}}].

\bibitem{PhysRevD.91.023514}
B.~J. Broy, D.~Roest, and A.~Westphal, {\it {Power Spectrum of Inflationary
  Attractors}},  {\em Phys.Rev.} {\bf D91} (2015), no.~2 023514,
  [\href{http://arxiv.org/abs/1408.5904}{{\tt arXiv:1408.5904}}].

\bibitem{BandM}
A.~L. Berkin and K.~Maeda, {\it {Effects of R3 and R Box R terms on R2
  inflation}},  {\em Physical Letters B} {\bf 245} (1990), no.~3 348.

\bibitem{Broy:2014xwa}
B.~J. Broy, F.~G. Pedro, and A.~Westphal, {\it {Disentangling the $f(R)$ -
  Duality}},  {\em JCAP} {\bf 1503} (2015), no.~03 029,
  [\href{http://arxiv.org/abs/1411.6010}{{\tt arXiv:1411.6010}}].

\end{thebibliography}\endgroup

\end{document}